\documentclass[fleqn,10pt,twocolumn]{wlscirep}
\usepackage{amssymb}
\usepackage{amsmath}
\usepackage{graphicx}
\usepackage{epsfig}

\title{Quantum phase transition induced by real-space topology}
\author[1]{C. Li}
\author[2]{G. Zhang}
\author[1]{S. Lin}
\author[1,*]{Z. Song}
\affil[1]{School of Physics, Nankai University, Tianjin 300071, China}
\affil[2]{College of Physics and Materials Science, Tianjin Normal University, Tianjin 300387, China}
\affil[*]{songtc@nankai.edu.cn}
\begin{abstract}
A quantum phase transition (QPT), including both topological and symmetry
breaking types, is usually induced by the change of global parameters, such
as external fields or global coupling constants. In this work, we
demonstrate the existence of QPT induced by the real-space topology of the
system. We investigate the groundstate properties of the tight-binding model
on a honeycomb lattice with the torus geometry based on exact results. It is
shown that the ground state experiences a second-order QPT, exhibiting the
scaling behavior, when the torus switches to a tube, which reveals the
connection between quantum phase and the real-space topology of the system.
\end{abstract}

\begin{document}

\maketitle

\flushbottom
%\affiliation{${\ }^1$School of Physics, Nankai University, Tianjin 300071, China \\
%${\ }^2$College of Physics and Materials Science, Tianjin Normal University,
%Tianjin 300387, China}

%%%{11.30.Er Charge conjugation, parity, time reversal, and other discrete symmetries}
%03.65.vf Phases: geometric; dynamic or topological
%03.65.-w Quantum mechanics

\section*{Introduction}

Quantum phase transitions (QPTs) are of central interest both in the fields
of condensed matter physics and quantum information. The transition
describes an abrupt change in the ground state of a many-body system due to
its quantum fluctuations.\ In general, a global physical parameter, such as
external fields or widely distributed coupling constants may drive QPTs,
including both topological \cite{Hasan} and symmetry breaking types \cite%
{SachdevBook}. During the transition, the real-space geometry of the system
is usually unchanged. A natural question is whether a change of real-space
topology can induce a QPT for some peculiar cases. It is traced back to a
problem in classical physics:\textbf{\ }the effect of a magnetic field on an
object depends on its real-space topology (see the illustration in Fig. \ref%
{fig1}). A magnetic field affects a conducting loop via\textbf{\ }the
magnetic flux, which is independent of its shape, for instance, no matter it
is a\textbf{\ }metal donut or cup. However, it has no effect on a metal bar.
The switch of the topology is equivalent to the sudden removal of the
applied magnetic field, which may introduce a sudden change of the quantum
state at zero temperature.

In this work, we demonstrate the influence of the real-space topology to the
quantum phase via concrete tight-binding models on a honeycomb and square
lattices, respectively. The change of the topology is presented by the value
of hopping constants, which connecting the two ends of a tube. When the
boundary hopping constants vary from finite to zero, a torus switches to a
tube. We investigate the groundstate property as the function of the
boundary coupling. Analytical and numerical results show that the ground
state exhibits the scaling behavior of second-order QPTs in the honeycomb
lattice, but not in the square lattice. It reveals that a geometric
topological transition may induce a QPT in certain systems, which display
physics beyond the current understanding of the QPT. The possible relation
between QPTs and the geometric quantity in real-space may open attractive
topics for different scientific communities.

\begin{figure}[tbp]
\centering
\includegraphics[ bb=41 653 557 800, width=0.45\textwidth, clip]{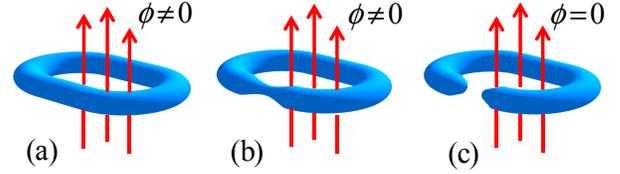}
\par
\caption{(color online). Schematic illustration of the aim of the present
work. It arises from the fact that the effect of a magnetic field on an
object depends on its real-space topology. We consider three cases: (a) A
perfect torus, (b) a cut torus, (c) a broken torus. Objects in (a) and (b)
have the same real-space topology, while (c) is topologically equivalent to
a bar, with different real-space topology. Charged particles in systems (a)
and (b) feel the same flux, while (c) cannot feel the existence of the
field. The switch of the topology is equivalent to the sudden removal of the
applied magnetic field, which may introduce a sudden change of the quantum
state at zero temperature.}
\label{fig1}
\end{figure}

\begin{figure}[tbp]
\centering
\includegraphics[ bb=42 440 547 791, width=0.45\textwidth, clip]{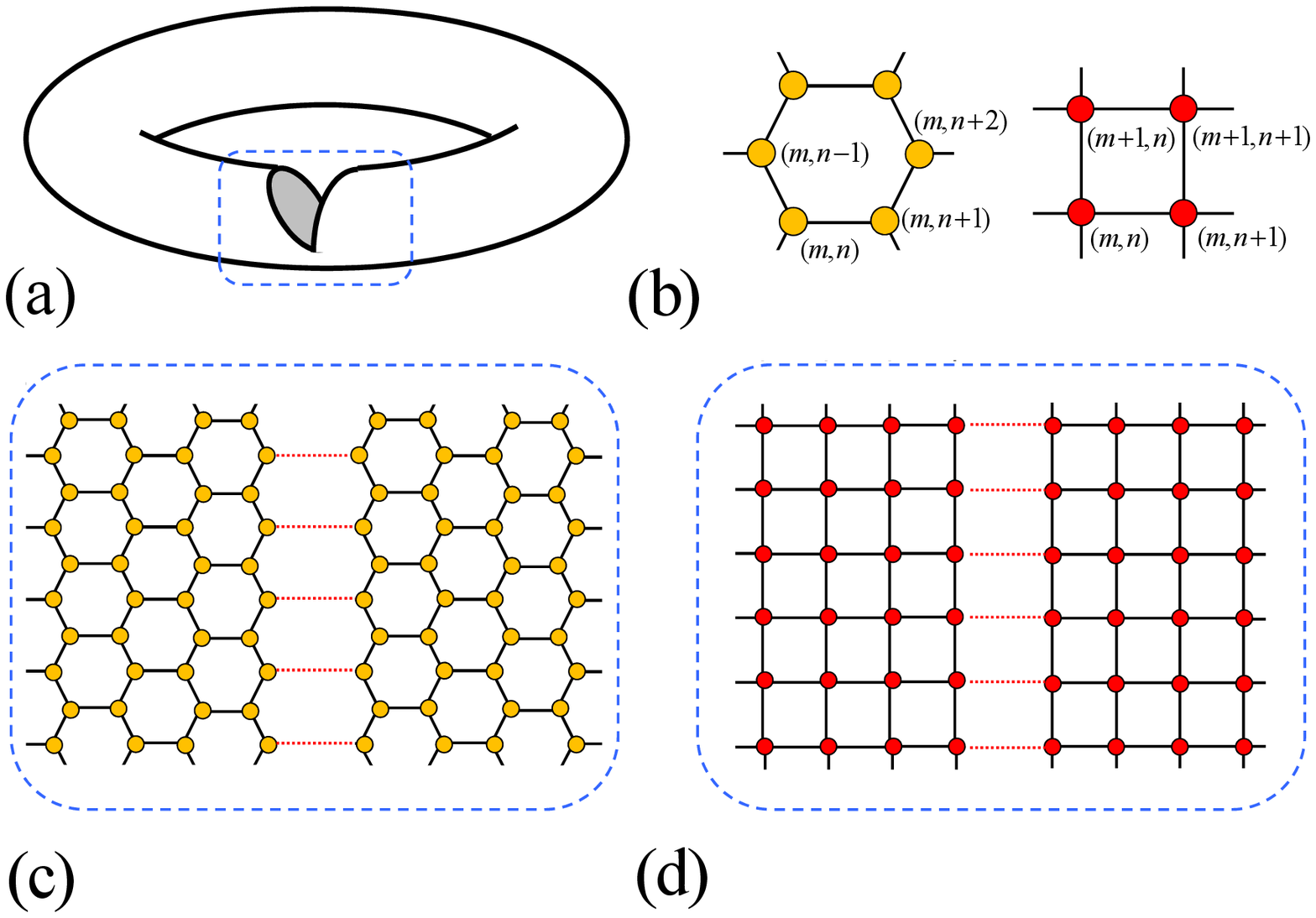}
\par
\caption{(color online). Schematic illustration of the lattice systems,
which are employed to construct the systems (a) with different real-space
topologies. We consider two types of lattices: (c) A honeycomb lattice; (d)
A square lattice. A cut or broken torus is presented by the weak or zero
hopping constants, which are indicated by dashed lines. The site index of
two types of lattices is indicated in (b).}
\label{fig2}
\end{figure}

\section*{Results}

\subsection*{Graphene torus}

We consider a system of noninteracting particles in a honeycomb geometry,
subjected to a magnetic flux $\phi $. The tight-binding model for this
system can be described by the Hamiltonian%
\begin{eqnarray}
&&H=-t\sum_{m=1}^{M}(\sum_{n=1}^{N-1}a_{m,n}^{\dagger
}a_{m,n+1}+\sum_{n=1}^{N/4}a_{m,4n}^{\dagger }a_{m+1,4n-1} \\
&&+\sum_{n=1}^{N/4}a_{m,4n-3}^{\dagger }a_{m+1,4n-2})-\eta te^{i\phi
}\sum_{m=1}^{M}a_{m,N}^{\dag }a_{m,1}+\mathrm{H.c.},  \notag
\end{eqnarray}%
where $a_{m,n}$ ($a_{m,n}^{\dagger }$) annihilates (creates) an electron in
site $\left( m,n\right) $ on an $M\times N$ lattice\ with integer $N/4$ and $%
M\geqslant 3$, and obeys the periodic boundary conditions, $%
a_{M+1,4n-1}=a_{1,4n-1}$ and $a_{M+1,4n-2}=a_{1,4n-2}$, with $m\in \left[ 1,M%
\right] $, $n\in \left[ 1,N/4\right] $. Parameter $t$ is hopping integral.
Here $\phi =2\pi \Phi /\Phi _{0}$, where $\Phi $\ is the flux threading the
ring, $\Phi _{0}$ is the flux quantum. In Fig. \ref{fig2}, the geometry of
the model is illustrated schematically. In this model, factor $\eta $\ is
important, determining the boundary condition of the system. In the view of
geometry, the value of $\eta $\ measures the topology of the system: nonzero
$\eta $\ corresponds to a torus\ whereas zero $\eta $\ stands for a bar. The
aim of this work is to explore what happens to the ground state when $\eta $%
\ passes the zero point. To this end, exact result is preferable. We note
that the value of $\eta $\ does not affect the translational symmetry in
another direction, so we employ the Fourier transformation%
\begin{equation}
c_{k,l}^{\dagger }=\frac{\xi _{l}}{\sqrt{M}}\sum_{j=1}^{M}e^{ikj}a_{j,l}^{%
\dagger },
\end{equation}%
to rewrite the Hamiltonian, where $\xi _{l}=1$ for $l=4n$, $4n-3$ and $\xi
_{l}=e^{-ik/2}$ for\ $l=4n-1$, $4n-2$ with $n\in \left[ 1,N/4\right] $, and $%
k=2\pi m/M$, $m\in \left[ 1,M\right] $. The Hamiltonian can be expressed as $%
H=\sum_{k}H_{k}$, where%
\begin{eqnarray}
H_{k} &=&-\lambda _{k}t\sum_{n=1}^{N/2}c_{k,2n-1}^{\dag
}c_{k,2n}-t\sum_{n=1}^{N/2-1}c_{k,2n}^{\dag }c_{k,2n+1}  \notag \\
&&-\eta te^{i\phi }c_{k,N}^{\dag }c_{k,1}+\mathrm{H.c.}.  \label{PR}
\end{eqnarray}%
Together with $\left[ H_{k},H_{k^{\prime }}\right] =0$, we find that $H$ is
a combination of $M$ independent Peierls rings with the $k$-dependent
hopping integral $\lambda _{k}=2\cos \left( k/2\right) .$

The one-dimensional dimerized Peierls system at half-filling, proposed by
Su, Schrieffer, and Heeger (SSH) to model polyacetylene \cite{Su,Schrieffer}%
, is the prototype of a topologically nontrivial band insulator with a
symmetry protected topological (SPT) phase \cite{Ryu,Wen}. In recent years,
extensive studies have been received \cite{Xiao,Hasan,Delplace,ChenS1,ChenS2}%
. For the open boundary condition, the number of zero modes reflects the
winding number as a topological invariant, according to the bulk-boundary
correspondence. Specifically, when $\eta =0,$ and infinite $N$, there are
two zero modes if\ $\left\vert \lambda _{k}\right\vert <1$, but not if $%
\left\vert \lambda _{k}\right\vert >1$. Accordingly, there are approximately
$M/3$ pairs of zero modes for a graphene tube with the open boundary
condition. It indicates that the groundstate energy changes arising from the
formation of the zero modes. We are interested in this process.

According to the Methods, we know that there is a pair of approximate
solution for the two zero modes of Eq. (\ref{PR}) for small $\eta $, which
are%
\begin{equation}
\mathbf{A}_{k}^{\pm }=\frac{1}{\sqrt{2}}(\mp \sqrt{\frac{\eta e^{i\phi
}-\lambda _{k}^{N}}{\eta e^{-i\phi }-\lambda _{k}^{N}}}\mathbf{A}_{k,0}^{+}+%
\mathbf{A}_{k,0}^{-})
\end{equation}%
with eigenvalues%
\begin{equation}
\varepsilon _{k}^{\pm }=\pm \Omega _{k}^{-1}t\sqrt{\left( \eta -\lambda
_{k}^{N}\cos \phi \right) ^{2}+\lambda _{k}^{2N}\sin ^{2}\phi }.
\label{epsilon_k}
\end{equation}%
Where%
\begin{equation}
(\mathbf{A}_{k,0}^{\pm })^{T}=\frac{1}{\sqrt{\Omega _{k}}}(\alpha
_{k,1}^{\pm },\alpha _{k,2}^{\pm },\alpha _{k,3}^{\pm },...,\alpha
_{k,N}^{\pm }),
\end{equation}%
with $\alpha _{k,l}^{+}=[1-\left( -1\right) ^{l}]\lambda _{k}^{\left(
l-1\right) /2}/2$, $\alpha _{k,l}^{-}=[1+\left( -1\right) ^{l}]\lambda
_{k}^{\left( N-l\right) /2}/2$\ and $\Omega _{k}=\left( 1-\lambda
_{k}^{2N}\right) /\left( 1-\lambda _{k}^{2}\right) $ $\approx \left(
1-\lambda _{k}^{2}\right) ^{-1}$. These analytical expressions are the base
for the investigation of the quantum phase transition induced by $\eta $.

\begin{figure*}[tbp]
\centering
\includegraphics[ bb=19 223 371 516, width=0.23\textwidth, clip]{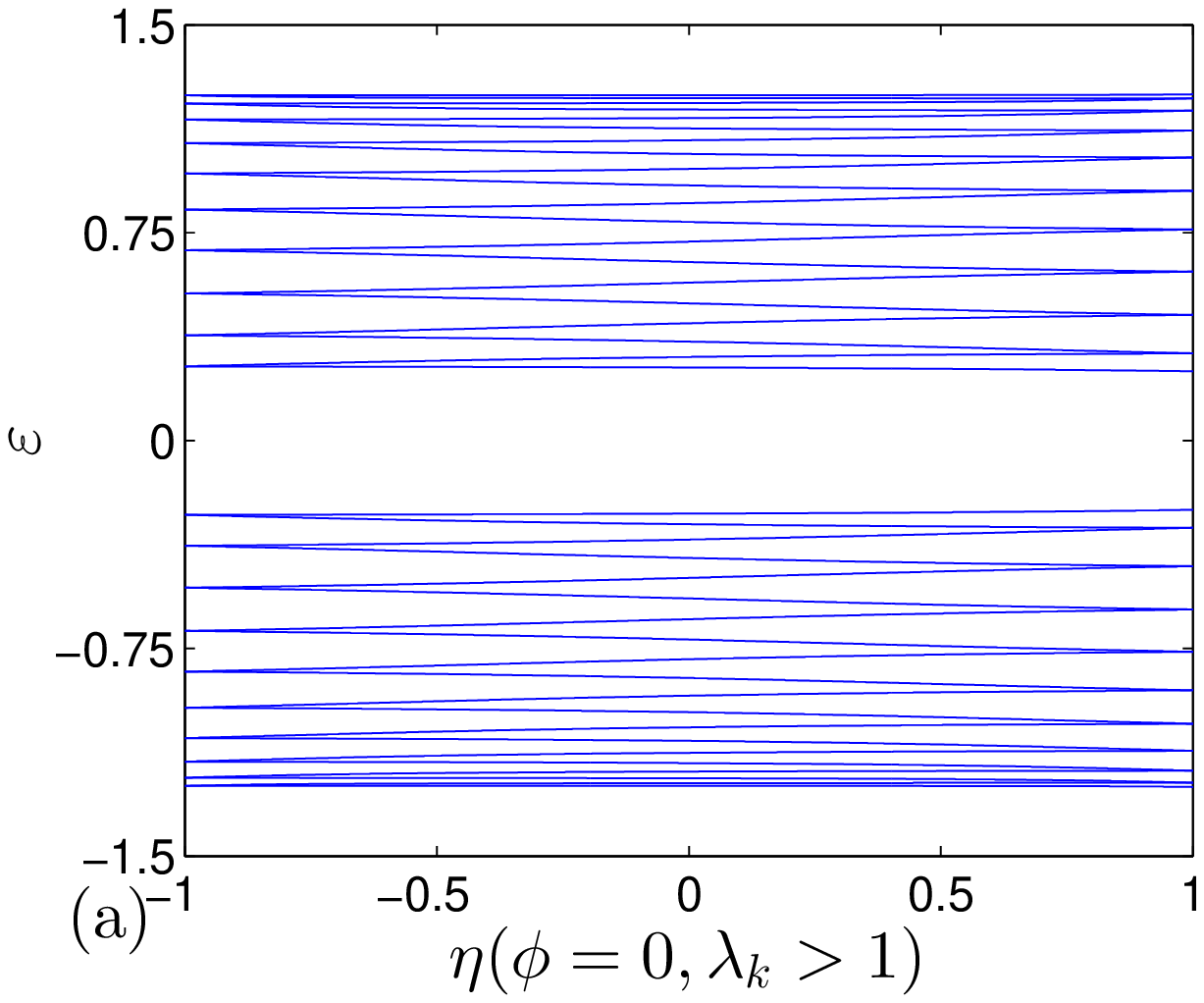} %
\includegraphics[ bb=19 223 371 516, width=0.23\textwidth, clip]{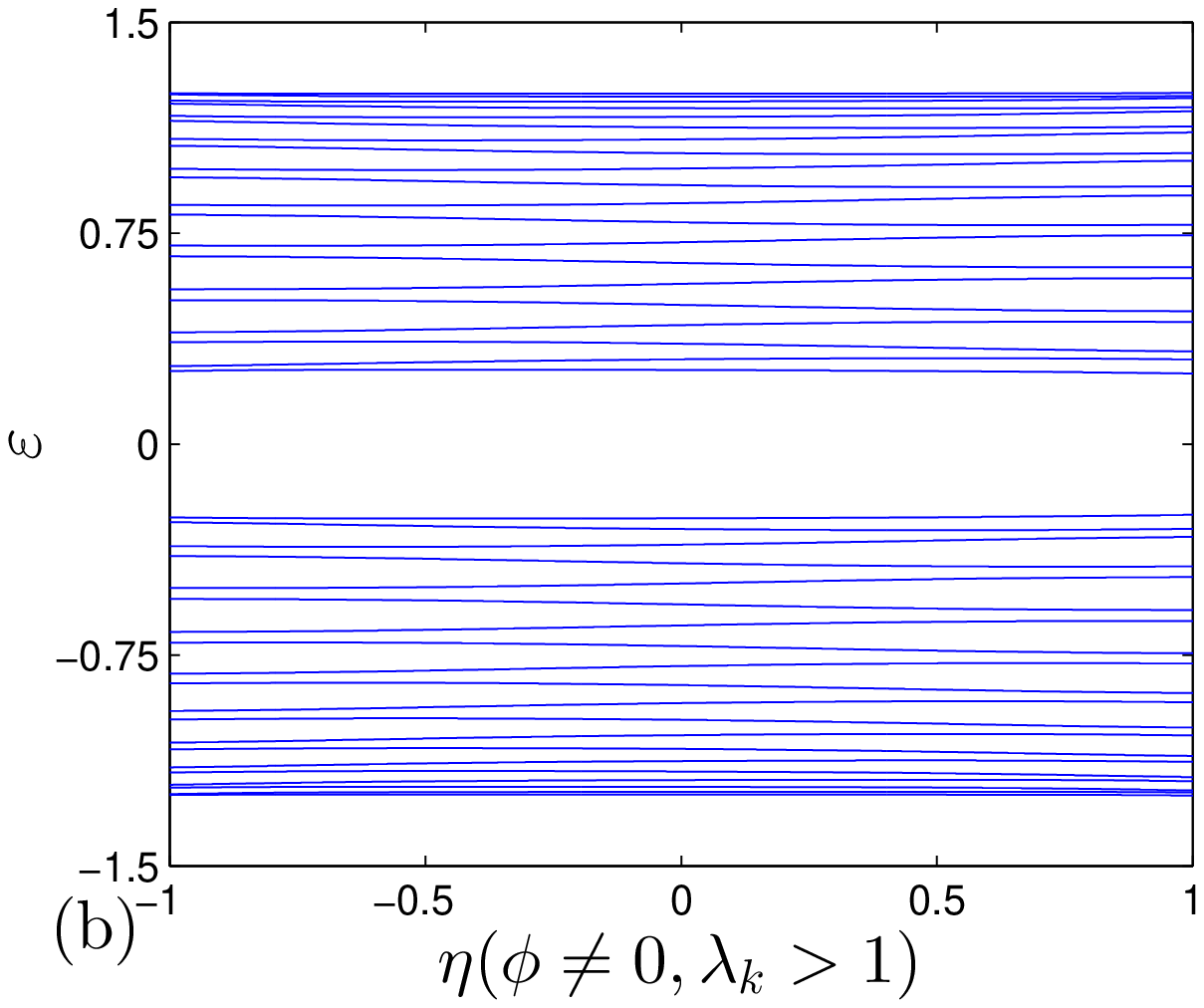} %
\includegraphics[ bb=19 223 371 516, width=0.23\textwidth, clip]{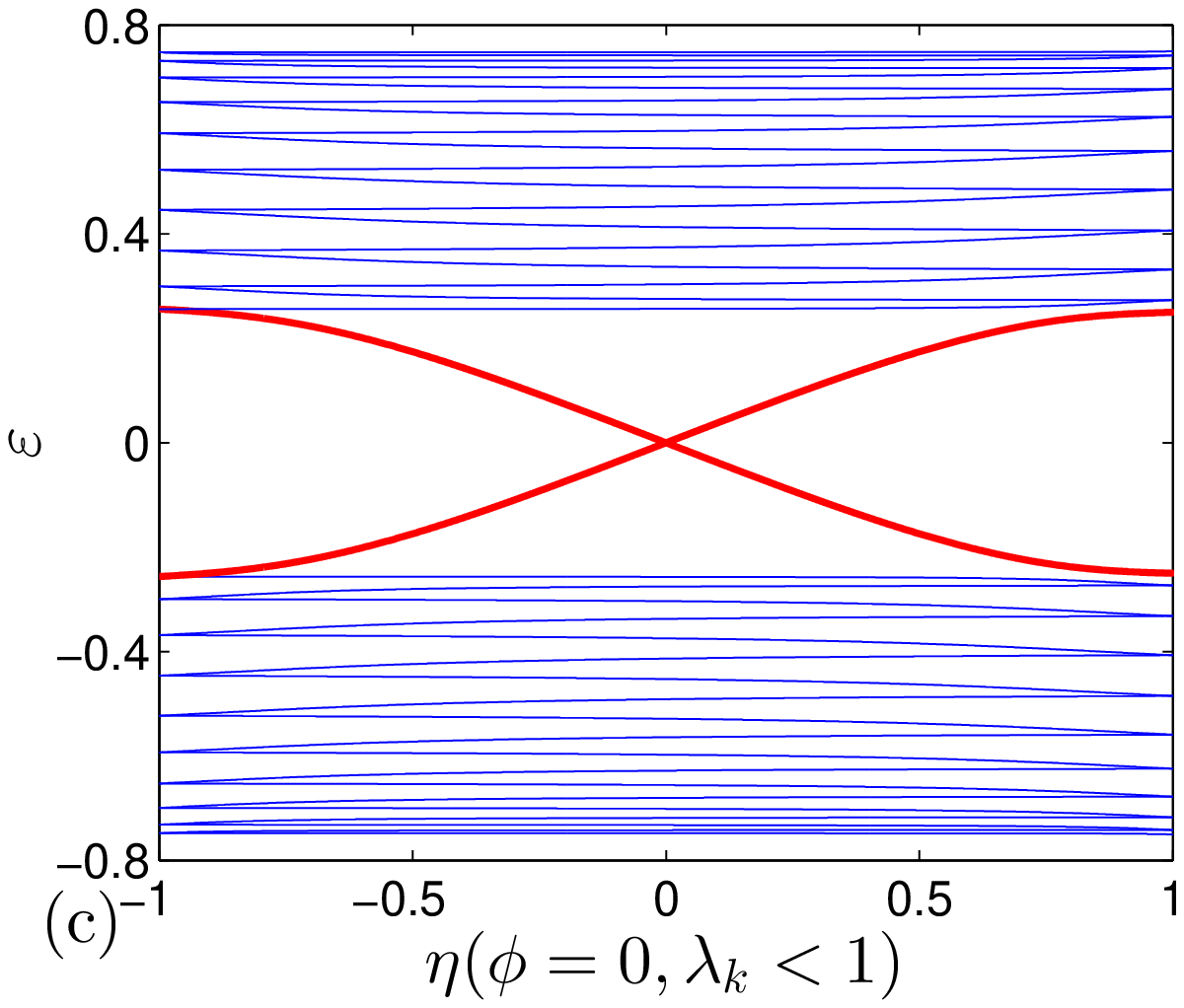} %
\includegraphics[ bb=19 223 371 516, width=0.23\textwidth, clip]{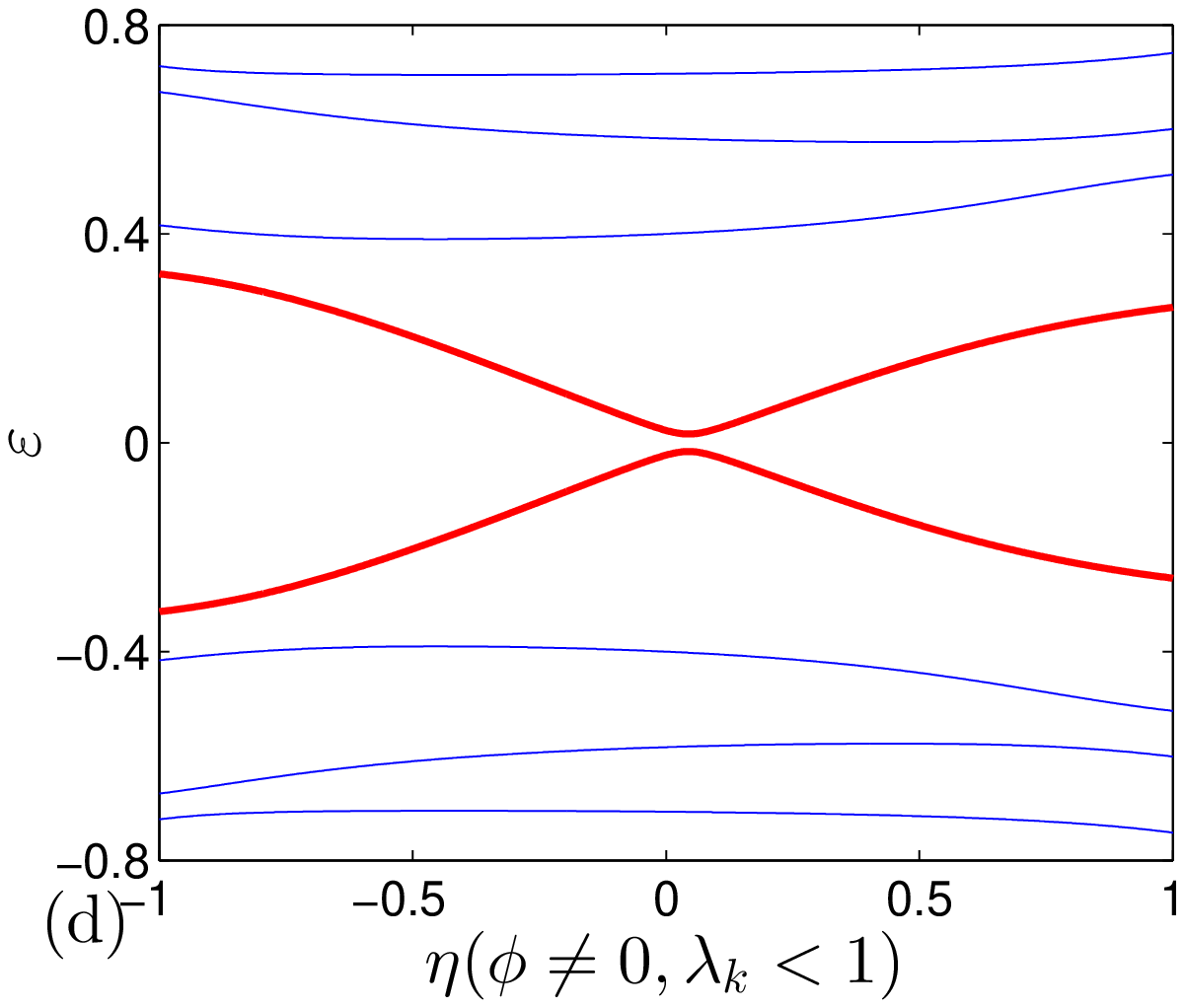}
\caption{(color online) Energy spectra for the Hamiltonian in Eq. (5) on a
lattice with $N=20$ for (a), (b), (c) and $N=4$ for (d), obtained by exact
diagonalization. The parameters are (a) $\protect\phi =0$, $\protect\lambda %
_{k}=1.5$; (b) $\protect\phi =\protect\pi /4$, $\protect\lambda _{k}=1.5$;
(c) $\protect\phi =0$, $\protect\lambda _{k}=0.5$; (d) $\protect\phi =%
\protect\pi /4$, $\protect\lambda _{k}=0.5$. We see that the zero modes do
not appear in the cases of (a) and (b), no matter the flux is zero or not.
In contrast, the zero modes appear as a level crossing and avoided level
crossing,\ in the cases of (c) and (d), respectively. In the case of (d), we
take a small $N$ in order to demonstrate the avoided level-crossing clearly.}
\label{fig3}
\end{figure*}

\begin{figure*}[tbp]
\centering
\includegraphics[ bb=17 229 366 500, width=0.32\textwidth, clip]{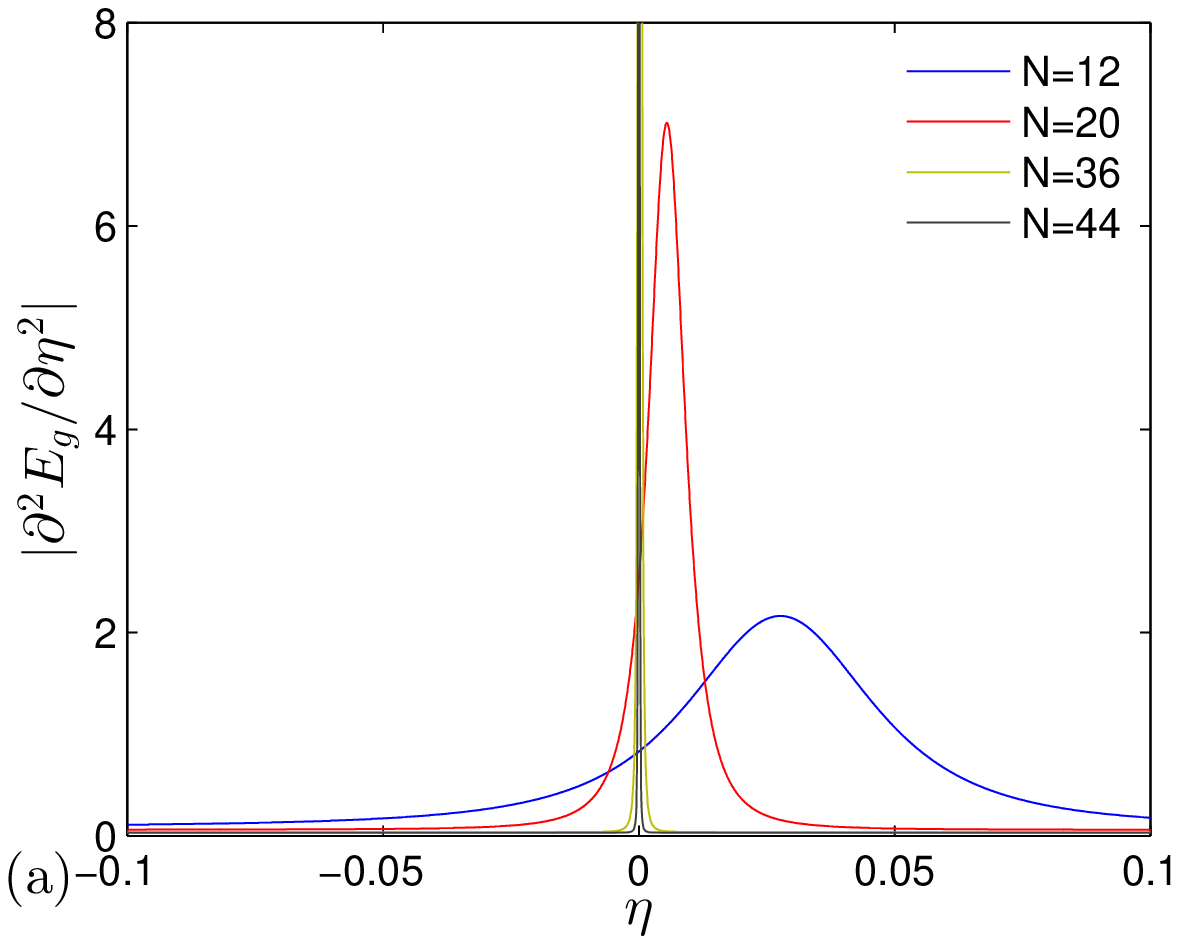} %
\includegraphics[ bb=17 229 366 500, width=0.32\textwidth, clip]{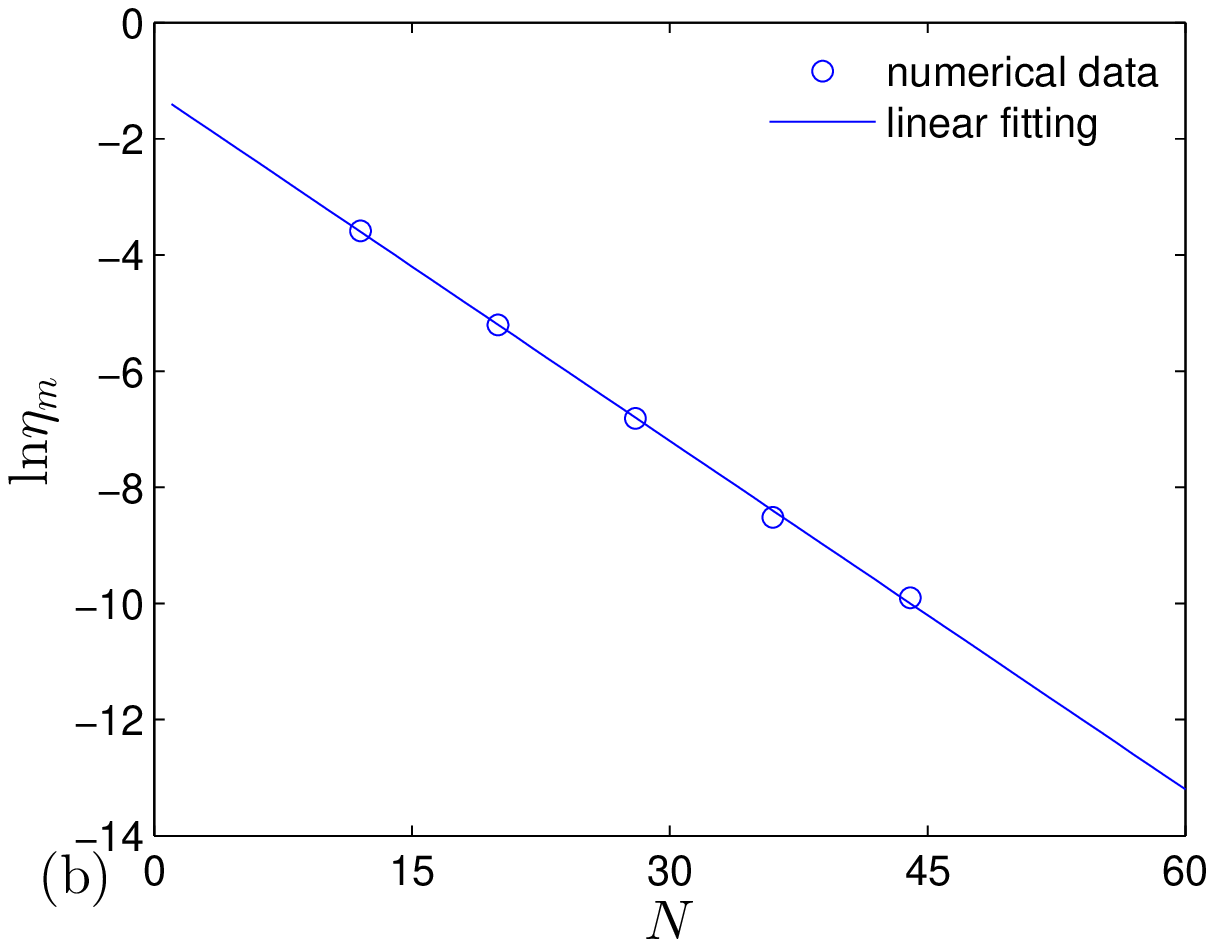} %
\includegraphics[ bb=17 229 366 500, width=0.32\textwidth, clip]{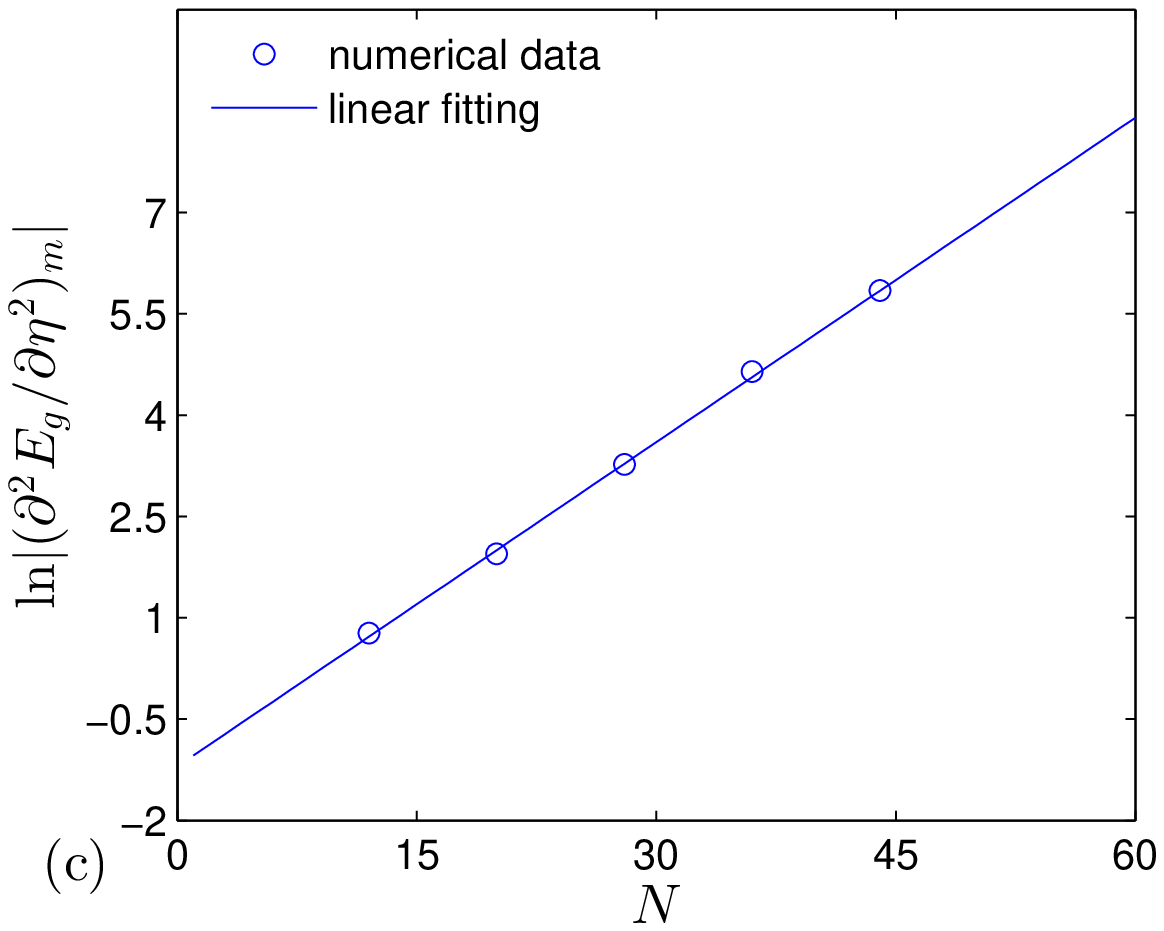}
\caption{(color online). The characteristics of second-order QPT for the
present system. (a) Plots of $\partial ^{2}E_{\mathrm{g}}/\partial \protect%
\eta ^{2}$ as a function of $\protect\eta $ for different values of $N$, (b)
the scaling law of pseudo critical point $\protect\eta _{m}$ as a function
of $N$, (c) the scaling law of the $(\partial ^{2}E_{\mathrm{g}}/\partial
\protect\eta ^{2})_{m}$ as a function of $N$. The parameters for all plots
are $M=7$ and $\protect\phi =\protect\pi /4$. }
\label{fig4}
\end{figure*}

\subsection*{Scaling behavior}

Based on the above analysis, the groundstate wavefunction can be expressed
as $\left\vert \psi _{\mathrm{g}}\right\rangle =\left\vert \psi _{\mathrm{b}%
}\right\rangle \left\vert \psi _{\mathrm{m}}\right\rangle $, with energy $E_{%
\mathrm{g}}=E_{\mathrm{b}}+E_{\mathrm{m}}$, where $\left\vert \psi _{\mathrm{%
b}}\right\rangle $\ is the lower band eigenstate with energy $E_{\mathrm{b}}$%
, and%
\begin{equation}
\left\vert \psi _{\mathrm{m}}\right\rangle =\prod_{\{k\}}\psi _{k}^{\dag }%
\mathbf{A}_{k}^{-}\left\vert 0\right\rangle   \label{midgap state}
\end{equation}%
is the midgap state with energy $E_{\mathrm{m}}=\sum_{\{k\}}\varepsilon
_{k}^{-}$, where $\{k\}$\ denotes the set of $k$ within the region $2\pi /3$
$<k<$ $4\pi /3$. In the thermodynamic limit, the band state is independent
of $\eta $, while the midgap state is dependent of $\eta $. To characterize
the quantum phase transition induced by $\eta $, we look at the second-order
derivatives of the groundstate energy%
\begin{equation}
\frac{\partial ^{2}E_{\mathrm{g}}}{\partial \eta ^{2}}\approx \frac{\partial
^{2}E_{\mathrm{m}}}{\partial \eta ^{2}}=t\sin ^{2}\phi \sum_{\{k\}}\frac{%
\lambda _{k}^{2N}}{\Omega _{k}^{4}\left( \varepsilon _{k}^{-}\right) ^{3}}.
\end{equation}%
Obviously, the property of $\partial ^{2}E_{\mathrm{g}}/\partial \eta ^{2}$\
depends on the behavior of $\partial ^{2}\varepsilon _{k}^{-}/\partial \eta
^{2}$. From the exact expression of $\varepsilon _{k}^{\pm }$ in Eq. (\ref%
{epsilon_k}), we note that the gap between $\varepsilon _{k}^{+}$\ and $%
\varepsilon _{k}^{-}$\ has a minimum
\begin{equation}
\Delta _{m}^{k}=2t\Omega _{k}^{-1}\lambda _{k}^{N}\left\vert \sin \phi
\right\vert
\end{equation}%
at $\eta _{m}^{k}=\lambda _{k}^{N}\cos \phi $, which is obtained from $%
\partial \varepsilon _{k}^{\pm }/\partial \eta =0$. In addition, $\partial
^{2}\varepsilon _{k}^{-}/\partial \eta ^{2}$ reaches the maximum
\begin{equation}
\left( \frac{\partial ^{2}\varepsilon _{k}^{-}}{\partial \eta ^{2}}\right)
_{m}=\frac{-t}{\lambda _{k}^{N}\Omega _{k}\left\vert \sin \phi \right\vert }=%
\frac{-2t^{2}}{(\Omega _{k})^{2}\Delta _{m}^{k}},
\end{equation}%
at the same point $\eta _{m}^{k}$. It is found that, quantities $\Delta
_{m}^{k}$, $\eta _{m}^{k}$, and $\left( \partial ^{2}\varepsilon
_{k}^{-}/\partial \eta ^{2}\right) _{m}$\ all exhibit scaling behavior.
These facts should result in the scaling behavior of $\partial ^{2}E_{%
\mathrm{g}}/\partial \eta ^{2}$. We would like to point out that, the above
analysis is not applicable to the situation of $k=\pi $. In this case, we
have $\lambda _{k}=0$, which reduces the SSH ring to a trivial case. On the
other hand, in the case of $\phi \rightarrow 0$, the avoided level-crossing
between $\varepsilon _{k}^{+}$\ and $\varepsilon _{k}^{-}$\ becomes a
level-crossing in a finite system. The second-order QPT becomes the
first-order one.

To demonstrate the origin of the critical behavior occurs in $H_{k}$, we
plot several typical band structures for $H_{k}$\ as functions of $\eta $ in
Fig. \ref{fig3}. It shows that in the case of $\left\vert \lambda
_{k}\right\vert >1$, the band structures are unchanged when the boundary
condition changes no matter the flux presents or not. In the case of $%
\left\vert \lambda _{k}\right\vert <1$, the main band structures are still
unchanged when the boundary condition changes. There are two midgap levels
emerge\ from\ the upper and lower bands, respectively. The appearance of
midgap levels does not depend on the flux. However, the key feature is that
the flux can lead to a level-crossing. The flux takes the role of quantum
fluctuation, driving the second-order QPT. We also plot the $\partial ^{2}E_{%
\mathrm{g}}/\partial \eta ^{2}$ as functions of $\eta $ and $N$ for nonzero $%
\phi $ in Fig. \ref{fig4}. It shows that $\partial ^{2}E_{\mathrm{g}%
}/\partial \eta ^{2}$ has a maximum $(\partial ^{2}E_{\mathrm{g}}/\partial
\eta ^{2})_{m}$\ at a pseudo critical point $\eta _{m}$. This predicts that,
in large $N$ limit, $(\partial ^{2}E_{\mathrm{g}}/\partial \eta ^{2})_{m}$
diverges at zero $\eta _{m}$. The plots of $\eta _{m}$\ and\ $(\partial
^{2}E_{\mathrm{g}}/\partial \eta ^{2})_{m}$ as functions of $N$\ indicate
the scaling law, exhibiting the second-order QPT behavior. The linear
fitting allows us to estimate the scaling functions as the from%
\begin{eqnarray}
\text{\textrm{ln}}\eta _{m} &=&-\frac{N}{5}-\frac{6}{5}, \\
\text{\textrm{ln}}\left\vert (\frac{\partial ^{2}E_{\mathrm{g}}}{\partial
\eta ^{2}})_{m}\right\vert &=&\frac{4N}{25}-\frac{6}{5},
\end{eqnarray}%
which are in good accordance with the numerical results.

Another way of looking at QPTs from the quantum information point of view is
ground-state fidelity \cite{Zanardi,Gu}. For our case, we focus on the
midgap-state fidelity, which is defined as

\begin{equation}
F\left( \eta ,\delta \right) =\left\vert \left\langle \mathbf{A}^{+}\left(
\eta -\delta \right) \right\vert \mathbf{A}^{+}\left( \eta +\delta \right)
\rangle \right\vert .
\end{equation}%
A straightforward derivation results in

\begin{equation}
F\left( \eta _{m},\delta \right) =\frac{\lambda _{k}^{N}\left\vert \sin \phi
\right\vert }{\sqrt{\delta ^{2}+\lambda _{k}^{2N}\sin ^{2}\phi }}\approx
\frac{\Omega _{k}\Delta _{m}^{k}}{2t\left\vert \delta \right\vert },
\end{equation}%
which also exhibits scaling law.

From above analytical and numerical analysis, we conclude that the
real-space topology can induce a QPT. Key features of such kind of phase
transition are that it is induced by a local parameter $\eta $, which is
similar to the impurity induced QPT \cite{Vojta}. We would also like to
point out that the flux is crucial for the transition. In the case of zero
flux, the quantum transition reduces to the first-order transition. The
parameter $\eta $\ drives the transition from one edge state\ to another
through the quantum fluctuation of the flux. Another point we want to
address is that the contexture of the toroid, honeycomb lattice, is also
crucial for the QPT. We demonstrate this by the following system.

In contrast to the graphene tube, we consider a system of noninteracting
particles in a square lattice, subjected to a magnetic flux $\phi $. The
tight-binding model for this system can be described by the Hamiltonian

\begin{eqnarray}
H &=&-t\sum_{m=1}^{M}(\sum_{n=1}^{N-1}a_{m,n}^{\dagger
}a_{m,n+1}+\sum_{n=1}^{N}a_{m,n}^{\dagger }a_{m+1,n})  \notag \\
&&-\eta te^{i\phi }\sum_{m=1}^{M}a_{m,N}^{\dag }a_{m,1}+\mathrm{H.c.,}
\label{SS}
\end{eqnarray}%
which obeys the periodic boundary conditions, $a_{M+1,n}=a_{1,n}$. From the
Methods, there is no QPT for the corresponding real-space topological change
in the square lattice system. These two examples indicate that the
occurrence of a real-space induced QPT strongly depends on the contexture of
the system. It also reveals a fact that the groundstate property must be
tightly connected to the topology of the system in which a real-space
induced QPT can happen.

\section*{Discussion}

In this work, we have demonstrated the existence of the QPT induced by the
real-space topology of the system. In contrast to the conventional QPT,
which is driven by a global physical parameter, such a QPT is induced by a
varying local parameter. Nevertheless, the characteristics of second-order
QPT, such as scaling behaviors of the second-order derivatives of
groundstate energy, pseudo critical point, and the fidelity of the
groundstate wavefunction, are all obtained. This finding reveals the
connection between the QPT and the real-space topology, which will motivate
further investigation.

\section*{Methods}

\subsection*{The approximate solution of a Peierls ring with the $k$%
-dependent hopping integral $\protect\lambda _{k}=2\cos \left( k/2\right) $}

We write down the Hamiltonian (\ref{PR}) in the basis $\psi _{k}^{\dag
}=(c_{k,1}^{\dag },$ $c_{k,2}^{\dag },$ $c_{k,3}^{\dag },$ $...,$ $%
c_{k,N}^{\dag })$ and see that%
\begin{equation}
H_{k}=-t\psi _{k}^{\dag }h_{k,N}\psi _{k},
\end{equation}%
where $h_{k,N}$\ represents a $N\times N$ matrix and contains two parts, $%
h_{k,N}=h_{k,N}^{0}+h_{k,N}^{\prime }$. Here two $N\times N$ matrices are%
\begin{equation}
h_{k,N}^{0}=\left(
\begin{array}{ccccccc}
0 & -\lambda _{k} & 0 & \cdots & 0 & 0 & \lambda _{k}^{N} \\
-\lambda _{k} & 0 & 1 & \cdots & 0 & 0 & 0 \\
0 & 1 & 0 & \cdots & 0 & 0 & 0 \\
\vdots & \vdots & \vdots & \ddots & \vdots & \vdots & \vdots \\
0 & 0 & 0 & \cdots & 0 & 1 & 0 \\
0 & 0 & 0 & \cdots & 1 & 0 & -\lambda _{k} \\
\lambda _{k}^{N} & 0 & 0 & \cdots & 0 & -\lambda _{k} & 0%
\end{array}%
\right) ,
\end{equation}%
and%
\begin{equation}
h_{k,N}^{\prime }=\left(
\begin{array}{ccc}
0 & \cdots & \eta e^{i\phi }-\lambda _{k}^{N} \\
\vdots & \ddots & \vdots \\
\eta e^{-i\phi }-\lambda _{k}^{N} & \cdots & 0%
\end{array}%
\right) ,
\end{equation}%
respectively. It is hard to get the explicit eigenfunctions of matrix $%
h_{k,N}$. Fortunately, we have two eigenvectors with zero eigenvalue, i.e., $%
h_{k,N}^{0}\mathbf{A}_{k,0}^{\pm }=0$, where

\begin{equation}
(\mathbf{A}_{k,0}^{\pm })^{T}=\frac{1}{\sqrt{\Omega _{k}}}(\alpha
_{k,1}^{\pm },\alpha _{k,2}^{\pm },\alpha _{k,3}^{\pm },...,\alpha
_{k,N}^{\pm }),
\end{equation}%
with $\alpha _{k,l}^{+}=[1-\left( -1\right) ^{l}]\lambda _{k}^{\left(
l-1\right) /2}/2$, $\alpha _{k,l}^{-}=[1+\left( -1\right) ^{l}]\lambda
_{k}^{\left( N-l\right) /2}/2$\ and $\Omega _{k}=\left( 1-\lambda
_{k}^{2N}\right) /\left( 1-\lambda _{k}^{2}\right) $ $\approx \left(
1-\lambda _{k}^{2}\right) ^{-1}$. We see that in large $N$ limit, $h_{k,N}$\
depicts an open chains: (i) $\left\vert \lambda _{k}\right\vert <1$, $\eta
_{k}$\ vanishes, a $N$-site ring becomes a $N$-site chain; (ii)\ $\left\vert
\lambda _{k}\right\vert >1$, $\eta _{k}$\ tends to infinity, a $N$-site ring
reduces to a $\left( N-2\right) $-site chain and a $2$-site separated dimer
with eigenvalues out of the bands. In both two cases, there are always two
zero-mode states, in which the particle probability locates around the
junction, which are so-called edge states.\ The solutions of both cases
accord with each other. The solution for $h_{k,N}$\ with finite $N$ is the
basement of the scaling behavior for the geometric topological transition.
Since two zero modes $\mathbf{A}_{0}^{\pm }$ are at midgap, $h_{k,N}$\ can
be regarded as a perturbation for small $\eta $. The degenerate perturbation
method gives the
\begin{equation}
\mathbf{A}_{k}^{\pm }=\frac{1}{\sqrt{2}}(\mp \sqrt{\frac{\eta e^{i\phi
}-\lambda _{k}^{N}}{\eta e^{-i\phi }-\lambda _{k}^{N}}}\mathbf{A}_{k,0}^{+}+%
\mathbf{A}_{k,0}^{-})
\end{equation}%
with eigenvalues%
\begin{equation}
\varepsilon _{k}^{\pm }=\pm \Omega _{k}^{-1}t\sqrt{\left( \eta -\lambda
_{k}^{N}\cos \phi \right) ^{2}+\lambda _{k}^{2N}\sin ^{2}\phi }.
\end{equation}

\subsection*{The exact solution of the square lattice system}

From the Hamiltonian (\ref{SS}), which obeys the periodic boundary
conditions, $a_{M+1,n}=a_{1,n}$. The geometry of this system is
schematically illustrated in Fig. \ref{fig2}(b, d). We employ the Fourier
transformation%
\begin{equation}
c_{k,l}^{\dagger }=\frac{1}{\sqrt{M}}\sum_{j=1}^{M}e^{ikj}a_{j,l}^{\dagger },
\end{equation}%
to rewrite the Hamiltonian, where $k=2\pi m/M$, $m\in \left[ 1,M\right] $.
The Hamiltonian can be still expressed as $H=\sum_{k}H_{k}$, where%
\begin{eqnarray}
H_{k} &=&-t\sum_{n=1}^{N-1}c_{k,n}^{\dag }c_{k,n+1}-\eta te^{i\phi
}c_{k,N}^{\dag }c_{k,1}  \notag \\
&&+\mathrm{H.c.}-\lambda _{2k}t\sum_{n=1}^{N}c_{k,n}^{\dag }c_{k,n}.
\end{eqnarray}%
Together with $\left[ H_{k},H_{k^{\prime }}\right] =0$ and $\lambda
_{2k}=2\cos k$, we find that $H$ is a combination of $M$ independent rings
with the $k$-dependent on-site potential $-\lambda _{2k}t$. The spectra of $%
H_{k}$\ with $\eta =1$\ and $0$\ are $-2t\cos [(2\pi n+\phi )/N]$ $-\lambda
_{2k}t$ and $-2t\cos [\pi n/(N+1)]$ $-\lambda _{2k}t$, respectively, where$\
n\in \lbrack 1,N]$. It indicates that the band structure is unchanged as the
boundary condition changes.

\section*{Acknowledgements}

We acknowledge the support of the National Basic Research Program (973
Program) of China under Grant No. 2012CB921900 and CNSF (Grant No. 11374163).

\section*{Author contributions statement}

C. L., G. Z. \& S. L. did the derivations and edited the manuscript. Z. S.
conceived the project and drafted the manuscript. All authors reviewed the
manuscript.

\section*{Additional information}

The authors do not have competing financial interests.

\end{document}